\documentclass[aps,pre,twocolumn,showpacs,epsfig,epsf,amssymb]{revtex4}
\usepackage{graphicx}
\usepackage{amssymb}
\usepackage{amsmath}
\usepackage{subfigure}
\usepackage{booktabs}
\usepackage{array}
\begin{document}

\title{Network formation and gelation in Telechelic Star Polymers 
}

\author{Indrajit Wadgaonkar$^1$, Apratim Chatterji$^2$ }
\email{apratim@iiserpune.ac.in}
\affiliation{
$1$ Flat I-308, Nakshatra I-Land, Moshi-Alandi Road, Pune-412105.\\ 
$2$ IISER-Pune, 900 NCL Innovation Park, Dr. Homi Bhaba Road,  Pune-411008, India.
}
\date{\today}
\begin{abstract}
We investigate the efficiency of gelation and network formation in telechelic star polymer melt,
 where the tips of polymer arms are dipoles while rest of the monomers  are 
uncharged. Our work is motivated by the experimental observations \cite{lele}, 
in which rheological studies  of telechelic star polymers of poly-(L-actide), a 
bio-degradable polymer, showed a drastic increase in elastic properties (up to $2000$ times) 
compared to corresponding  star polymers without the telechelic arm ends. 
In contrast to previous studies, we avoid using effective attractive Lennard Jones potentials or 
dipolar potentials to model telechelic interactions.
Instead we use explicit Coulomb positive and negative charges at the tip of polymer-arms of our bead-spring model 
of star polymers. By our simulations we show that the dipoles 
at the tip of star arms aggregate together to form clusters of dipoles. Each cluster has 
contribution from several stars, and in turn each star contributes to several clusters. Thus the entire
polymer melt forms a connected network. Network forming tendencies decrease with 
decrease of the value of the effective charge constituting the dipole: this can be experimentally realized by 
choosing a different ionomer for the star tip. We systematically varied the value of dipole 
charges, the fraction of star-arms with dipoles at the tip and the length of the arms. 
The choice of explicit charges  in our calculations enables us to make 
better quantitative predictions about the onset of gelation, moreover we get qualitatively
distinct results about structural organization of dipoles within a  dipole-cluster. 
\end{abstract}

\maketitle
\section{Introduction}

The need for designable and tunable biodegradable polymers cannot be overemphasized in the present scenario.
Often the synthesized biodegradable polymers do not have the required properties, and then suitable modifications
have to be implemented on the polymer chains to get the desired properties \cite{meta0,meta1,meta2,dolog}. 
One such example of a synthesized 
polymer melt has been poly-(L-lactide), a biodegradable and bio-renewable polymer \cite{lele}. But unfortunately,
the melt strength,  the maximum tension that can be applied to the melt without breaking, of poly-lactide 
is quite low which makes it unsuitable for  extrusion to thin plastic sheets or pipes or bags.  Melt strength,  
an indication of the value of the elastic response modulus $G^{'}$ of the melt, increases with the decrease 
of the viscosity $\eta$ of the melt. It is obvious without increase of $G^{'}$, poly-(L-lactide) cannot be used as 
a replacement for standard polymeric products available for use.

Using some ingenious chemistry,  $1-4\%$ of the L-lactide monomers were replaced by another 
suitable ionic group \cite{lele}, the  elastic modulus of poly-(L-lactide) was increased by a factor of $2000$ times.
The question is how and why is that possible? In this paper we investigate  the 
microscopic  basis of this huge increase in the elastic modulus  using simple bead spring models 
of polymers and properties of telechelic polymers. We quantify limits to which the experimentally suggested method 
of increasing the $G^{'}$ \cite{lele} or the melt strength can be explored and extended.

As per experimental evidences \cite{lele}, suitably polymerized L-lactic acid (PLA) is a star polymer with 
$6$ arms with $25$ lactic acid monomers per arm. Thus a PLA star has 150 monomers, and the elastic and viscous 
response functions, $G^{'}$ and $G^{''}$ were measured to be $1$ and $10$ pascals, respectively, in the newtonian rheology regime.  
When the monomers of the $6$ arms of stars polymers were suitably substituted to have $Na^{+}-COO^{-}$ ionomers
only at the tip of star arms, the $G^{'}$ increased to $2000$ pascals and the $G^{''}$ to $500$ pascals. When further 
rheological experiments with different number of ionomers per star were performed, the elastic 
response increased by a factor varying from $10$ to $2000$ times, as the number of ionic end groups per star $f$ 
was varied. Experimental control can be achieved such that the star melt has on an average 
just $2$ or $3$ or $4$ ionic end groups
per star. The reader is encouraged to appreciate that the change in composition is in just $2$,$3$ or $6$ monomers out 
of $150$ monomers in a star and just at arm-tips, but the increase in elastic and viscous response is huge. 

There are been previous work as well on linear and star polymers with different functionalized end groups at the tip 
of a polymer chains \cite{fetters,vlasso}, and they observe an increase of visco-elastic response of polymers 
depending on the nature of ions and architecture of polymers. The general expectation and understanding is that 
the ionomers form clusters of telechelic sections of chains, and this ends up in the physical gelation
of polymers chains \cite{vlasso}. This could lead to formation of star polymers starting from telechelic linear polymers, or induce 
conformational changes in individual star-polymers \cite{likos,likos1,blaak2} or in the structural rearrangements in the 
large scale organization of stars \cite{likos2,blaak}. 
Other theoretical/computational studies with telechelic chains have focussed on 
finding the sol-gel phase diagram of telechelic polymers in dilute polymeric systems  or the dynamical 
properties of associating polymers due to telechelic ends \cite{sanat,sanat1,berthier} including change in glass 
transition temperature \cite{eisenberg}. But nearly all studies telechelic polymers, also known as end-functionalized
polymers in literature \cite{roland}, model the attraction between telechelic ends by an attractive Lennard 
Jones  potential with a cut off at a suitable distance. Moreover, most of the the theoretical studies of 
telechelic stars stick to the dilute limit.
Previous experimental studies using poly-(L-lactic acid) ionomers had considered linear polymers 
and observed the increase in glass transition $T_g$ due to the presence of ionomers \cite{weiss1,weiss2}.

In a departure from previous computational studies, in our study we focus on a bead-spring model of a 
star polymer {\em melt}  with telechelic ends modelled as dipoles with {\em explicit charges} instead of 
effective attractive potentials.  The usually used effective attractive potentials (e.g. Lennard Jones)
used to model telechelic properties provide attraction at short length scales, 
on the contrary the Coulomb interaction acts between monomers far separated in space. 
We consider stars with $6$ arms, and $25$ monomers per arm in tune with the experiments \cite{lele} which motivated 
this investigation. The last two monomers of the star arms are replaced by a postively and negatively charged 
monomer, such that we have a dipole which in turn attempts to model the presence of $Na^+$ and $COO^-$ ionomers
at the arms tips of poly-(L-lactide) stars. We carry out Molecular dynamics simulation of such star ionomers, and vary the 
values of effective charges $\pm qe$ at the star polymer ends, where $q$ is a fraction $<1$ and $e$ is the electronic 
charge. Variation of the value of $q$ at the star tips would experimentally correspond to substituting different ionomers
at the tip of star arms, as has been considered in a previous study \cite{vlasso}. 

We establish at which values of $q$ do multiple dipoles 
aggregate together to form dipole-clusters overcoming thermal effects. These dipole clusters are multiply
connected to many stars, and each star contributes to many dipole clusters thereby forming a gel-like interconnected 
network of polymers. If one has just 2 or 3 dipoles per star, then obviously the dipole clusters formed are smaller
and then one has macromolecular assemblies instead of system-spanning percolating networks of stars connected through dipole
clusters. We do not compute dynamical quantities like viscosity or $G^{'}$ in our simulations as the calculations are 
too expensive, instead we focus on morphological quantities and deduce that the relaxation times will increase per
microscropic structure changes.

We emphasize that in contrast to previous studies of dipolar fluids \cite{dipole1,dipole2,dipole3,dipole4,dipole5,dipole6}, 
where the authors have used the dipolar potential as a $1/r^2$ potential valid at large distances away from the dipoles, 
we use explicit charges $\pm qe$ to model the dipoles. In our work the interaction energy between the dipoles is thereby 
calculated using explicit Coulomb potential between each pair of charges. This is necessary because the dipolar
monomer pairs can be atomistically close to each other where the multipole Taylor expansion is not even valid. As a consequence,
the structure and and organization of dipoles in a dipole cluster in our studies is different from what has been found 
in previous investigations of dipolar fluids \cite{dipole4,dipole5,dipole6}
 when the interaction potential between dipoles is modelled as  $1/r^2$ potential.  

The rest of the paper is organized as usual, the next section consists a detailed description of our model where 
we have taken extra effort to connect  with experimental numbers; we also state when we deviate from
experimental conditions. We do not use the effective charges $q$ for $Na^+$ and $COO^-$, instead vary it as a parameter.
In section III, we present our results, and finally we conclude in section IV with a discussion and future outlook.

\section{Method} 
Molecular dynamics simulations of PLA star ionomers were performed using a 
bead spring model of polymers. Each lactic acid monomer was modeled as a spherical bead connected by harmonic 
springs to the neighbouring monomers. The spring interaction between two neighbouring monomers are given by
\begin{equation}
V_{spring} = \kappa (x- \ell_0)^2
\end{equation}

where $\ell_0$ is the mean distance between monomers and $\ell_0 =1$ sets the length scale of the simulation.
Each star polymer has $6$ arms  and there are $L=25$ lactic acid monomers per star. The six linear polymer arms are 
attached to a central sphere by 
\begin{equation}
V_{sphere} = \kappa_s (x- \ell_0^{sph})^2
\end{equation}
where $\kappa_s = \kappa = 1000 k_B T$ and $\ell_0^{sph} = 2 \ell_0$. We set $k_B T =1$ and all other energies 
are measured in units of $k_BT$, e.g., $\kappa = 1000 k_BT$. A very high value of  $\kappa = 1000 k_BT$ 
is chosen to render the polymer arm to be  inextensible chain. Excluded volume interaction between 
the beads are incorporated by a suitably shifted purely repulsive Lennard Jones interaction truncated 
at a distance $r_c = 2^{1/6} \sigma $, where $\sigma =0.8 \ell_0$  is the diameter of the monomeric beads. 
The excluded volume radius of the central sphere is $ \ell_0^{sph} - 0.5\sigma =1.6 \ell_0$. We have chosen a large central
sphere so that we can add more number of arms around central sphere in future studies as in 
\cite{ac1,ac2,ac3}. The mass $M$ 
of each monomer is set as $M=1$. Giving suitable values to $k_BT$, $M$ and $\ell_0$, we can calculate 
suitable values to quantitites of interest like $\kappa$, the calculated average energy of the system 
or unit of time $\tau= \sqrt{(M\ell_0^2/k_BT)}$.  For example, setting $T=300$K, $\ell_0 =1nm$ and 
$M=15 \times 10^{-23} kg$ for a lactic acid monomer, we get $\tau=0.2$ nano-seconds.  But for purpose 
of simulation, we set $k_BT=1,M=1,\ell_0=1$ and measure all other quantities in these units.

The presence of ${\rm Na^{+}}$ and ${\rm COO^{-}}$ at the end of the arms of the PLA star results 
in an effective dipole at the tip of every polymer arm; these dipoles in turn interact and 
attract/repel each other depending upon their relative orientation. We do not model dipolar 
interaction by the $\frac{1}{r^2}$ dipolar potential as this approximation breaks 
down at distances when the dipoles are close to each other, instead  we use the Coulomb potential
between each pair of charges. We consider the 25-th and 24-th monomer at the tip  of each arm 
to have a charge of $+qe$ and $-qe$, respectively, where $q < 1$ is a fraction and $e$ is the electronic
charge.  In our simulations, we can set all the $6$ arms to have dipoles at the tip. 
 In this case, the number $f$ of dipoles per star is $f=6$. Alternatively, we can choose to 
have a system where only  $2$ or $3$  out of $6$ arms  have charges at the tips of polymer arms, 
which then corresponds to $f=2$ or $f=3$, respectively.  To compare our simulations results to 
experiments, ideally we would also need to know the effective charges  on the ${\rm Na^+ }$ or 
${\rm COO^-}$ ions at the end of the PLA arm. Since we do not know the effective
charge $q e$ at each ionomer, we use 4 different values of effective charges $qe$  
and analyze the network formation between stars for each value of $q$. 

To put numbers into perspective, two isolated electronic charges  $+e$ and $-e$  at a distance of $\ell_0 =1$ nm from each other 
have a Coulomb energy $E_c(e,1nm) \approx 61 k_BT_1$ for $T_1=300K$.  For  our simulations we use 4 different values of 
effective charge  with appropriate value of $q$  such that $ R=E_c(qe,\ell_0)/k_BT$ has values $R=5,10,20,40$,
respectively. Since $E_c(qe,1nm) = q^2 61 k_BT$, the value of $R$ can also be expressed as $R=61 q^2$, 
and as before $R$ remains dimensionless.  We then study how $R$ affects  structural arrangement of stars and 
dipoles at the microscopic length-scale.  The calculation of Coulomb forces  in a finite box with periodic
boundary conditions (PBC) would necessitate  the use of Ewald summation techniques or alternatively 
$P^3M$ (particle-particle particle mesh),
especially since electrostatic interactions have long ranged $\sim 1/r$  potentials. We use LAMMPS simulation 
package \cite{lammps} with a cubic box. LAMMPS has inbuilt $P^3M$ \cite{deserno,deserno2} implemented 
which we use to calculate Coulomb interaction between dipoles at tips of arms. 

For our simulations we maintain our star polymer melt densities close to that used in experiments \cite{lele}. The 
PLA melt density of stars in 1.06 ${\rm gm/cm^3}$ and molecular weight of stars is $10000$ gm/mol. Thereby 
 $ 1 {\rm cm^3}$ contains $ \sim 10^{-4}$ moles of PLA stars $ \sim 6 \times 10^{19}$ stars, and thus we calculate that 
one star occupies $16.7 \times 10^{-27} m^3$. If we assume that the star is a sphere occupying the specified 
volume,  we can estimate the approximate value of star-radius to be $\approx 16$ \AA. To the first approximation, the radius 
of the star is equal to the radius of gyration  $r_G$ of a polymer arm, thus 
$ r_G = = L^{0.6} l_0^e/\sqrt{6} =16 \AA$ , where $L=25$ is the number of monomers in a star arm. 
So the effective bond length $\ell_0^e$ is  $5.6$ \AA, which is nearly half the length $\ell_0 =1nm$ used to 
estimate $R$ in our simulations,
So a simulation box of $50 \times 50 \times 50 (\ell_0^e)^3$ should have $\sim 1300$ star polymers. 

\begin{figure}[!htb]
\centering
\includegraphics[width=0.75\columnwidth,angle=0]{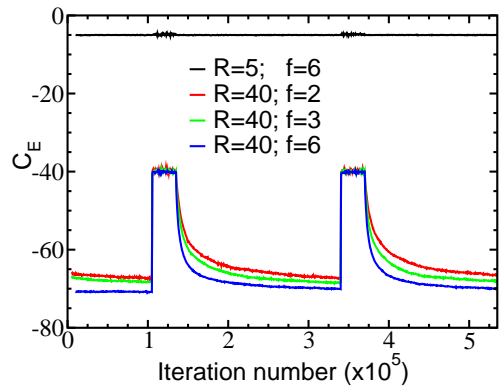} 
\caption{\label{energy} The average Coulomb energy per dipole $C_E$ versus the number of iterations over 
$2$ heating and cooling cycles starting just after the first $10^5$ iterations of an equilibrated system.
The system is heated to $50 k_BT$ for $30000$ iterations, and then 
allowed to equilibrate for next $10^5$ iterations at $1k_BT$, and we then collect statistical data for the 
next $10^5$ iterations.  Data is shown for $R=40$ and $R=5$. The fluctuations in $C_E$ is more
at higher temperatures.
}
\end{figure}

For our simulations we deviate from exact experimental values to study a less dense system. For computational
ease, we take $350$ stars in a cubic box of volume 
$L_{box}^3 = 50 \times 50 \times 50 \ell_0^3$, and carry out our simulations to investigate the equilibrium 
structure of the stars and the resultant clustering of dipoles  as a function of $R,f$ for $L=25$ and 
$L=50$. Different values of $R$ correspond to different values of partial
charge $qe$ on the charges constituting the dipoles at polymer arm tips. 
For our simulations we start with the equilibrated configuration of a single star polymer 
(without dipoles at arm tips) in a simulation box, then 349 copies of this equlibrated configuration are placed 
and packed in a lattice within a  $50 \times 50 \times 50 \ell_0^3$ simulation box and then equilibrated 
for $10^5$ iterations using  Molecular dynamics (MD) to create a melt of star polymers. 
Then dipolar charges at the tips of arms are switched on and the system is further equilibrated 
($10^5$ iterations) to have a melt of star polymers  with dipoles at arm tips. 
During equilibration we use a thermostat which rescales the velocities every $20$ iterations to maintain temperature 
$k_BT =1$. Integration time step  was chosen to be $\delta t = 0.001 \sqrt{(M \ell_0^2/k_BT)} =0.001 \tau$.  
Independent runs were given to cross-check that the star ionomer melt reached the same equilibrium energy and
structural arrangement of stars. 
Note that the number of arms per star always remains fixed at $6$; when $f=2$ it implies that only $2$ out of 
$6$ arms have dipoles at the tips of arms while 4 star-arms remain uncharged.

Because a melt of gel-like polymers is a dense system with long relaxation times, we have to ensure 
that the statistical quantitities that we measure are not the properties of a configuration stuck
in a initial condition dependent free energy minimum. To that end, after completion of equilibration 
and collecting statistical data for the initial $10^5$ iterations, we heat the system to $T_{50}= 50 T$ 
(where $k_BT=1$) and keep the system 
at high temperatures for $30000$ iterations, such that the system gets thermalized at high $T$.
Then the system is cooled down back to $T$, allowed to equilibrate for $1.05 \times 10^5$ 
iterations. We checked  that a completely different and statistically independent configuration
 and dipole cluster is formed after the cooling.  Then statistical data is collected over the next 
$10^5$ iterations, every $200$ steps before heating it again to $50k_BT$.  
This heating and cooling cycle is shown in Fig.\ref{energy} where the energy per dipole $C_E$ is shown
for different values of $f$ at $R=40$ for $L=25$ stars. The value of $C_E$ shoots up when temperature
is $T_{50}$, but then relaxes to equilibrium in around $10^5$ iterations once the temperature is 
reset to $T$.  

The heating to $T_{50}=50 T$ and cooling was carried out $4$ times, the statistical data was compared 
and seen to be equivalent in each of $5$ sets of runs over which data was collected.  For example, 
the average number of clusters in the box  for the system $R=40,f=6 $ and $L=25$ 
in the 5 individual runs were $217,210,216,198$ and $202$, respectively.  
Similarly, the average number of clusters  for the system  $R=5,f=6 $ and $L=25$ in the 5 set of runs were
$1701,1774,1775,1779$ and $1774$, respectively. We also compared the mean size of dipole-aggregates 
formed across $5$ runs and found them to be comparable within statistical fluctuations.
The data that we present in the results section is a statistical average of the initial 
run and $4$ runs after heating-cooling and equilibration.  Though $30000$ iterations at $50 k_BT$ 
might not result in the diffusion of stars over length scales comparable to  the diameter of the 
stars, it is enough to break any dipole clusters and make the arms move considerably in phase 
space. To that end, after cooling the dipole cluster configuration is completely 
independent of previous configuration. 

In Fig.\ref{energy}, we show the average Coulomb energy per dipole $C_E$ versus number of iterations
with $R=40$ for $3$ different values of $f$ (and hence different dipoles densities), as well as  for 
$R=5$ for $f=6$.  After the initial equilibration of the system at $k_BT=1$, the Coulomb energy per dipole 
$C_E$ for $R=40, f=2$ system relaxes to  $\approx 65 k_BT$. 
This is lower than the energy of $2$ charges kept at a unit distance from each other, which is $40 k_BT$.  
Presumably dipoles attract and come together to form dipole aggregates and each charge interacts with many other
charges.  But $C_E$ goes up to $40 k_BT$ when the temperature is increased to $T_{50}$ (such $T_{50}/T =50$) 
at around $10^5$ iterations.   We checked thermal energy  at $T_{50}$ disintegrates any dipole clusters.  
For $f=3$ the value of $C_E(T)$ is slightly lower than that for $f=2$ stars, and for $f=6$ with 
higher density of charges $C_E \approx 70 k_BT$.  
The value of $C_E/k_BT$ goes nearly to $40$ at temperature $T_{50}$ for all values of $f$. 
This is not difficult to understand, as the thermally averaged effective interaction between 
freely rotating dipoles become a effective short range attractive $1/r^6$ interaction 
as is well discussed in the classic book by Israelachvili \cite{isra}.



The primary message from Fig.\ref{energy} is that  when the temperature is reduced 
back to $T_1$ from $T_{50}$, the Coulomb energy relaxes back to the same lower
value in approximately $ 10^5$ iterations for each value of $f$. This assures us
the system is equilibrated and we can start collecting data for statistical averages after this cooling step
till the system temperature is again hiked to $T_{50}$.
We also checked the value of $C_E$ for $R=5$ for $f=6$, the Coulomb interaction energy is nearly $5$ 
even at $k_BT=1$; though there is more energy fluctuations when the temperature is hiked to $T_{50}$.  This 
would imply the organization of dipoles at low and high temperature is similar and thermal energies 
overwhelms Coumlomb correlations or clustering.
Indeed we show later that for $R=40$ we get clusters of dipoles,  whereas for $R=5$ most of the dipoles
do not form aggregates with other dipoles.

\section{Results}
In this section we discuss our measurements and conclusions regarding the structure formation in 
telechelic star polymers in some detail.
We ask and quantify $4$ primary questions:
\begin{enumerate}
\item Do the dipoles at the tip of star-polymer arms aggregate together to form clusters of dipoles? 
If yes, how big are the clusters?, i.e., How many dipoles are there in one cluster? If many clusters 
are formed, what is the distribution of the cluster sizes? These set of questions are analyzed and 
quantified in Figs. \ref{total},\ref{fig1},\ref{fig1a} and \ref{fig2}.  
\item Do the dipoles of a particular cluster belong to a single star?
 Or does a dipole-cluster have contributions from the arms from different star polymers?  If so, how many stars?

A dipole cluster can behave as a node at which different star arms
get attached and are held together due to dipolar attraction. If some of the clusters have more than (say)
$6$ or $10$ dipoles (or more), and if most of the clusters have contributions
from many  stars, then each dipole-cluster connects up many star-polymers. Then the likely scenario 
is that all the stars arms in the system
will form a percolating network connected by dipole clusters and the system would be akin to 
gel state of polymers. These set of questions are investigated using data presented  in Fig. \ref{fig3} and \ref{fig4}.
\item Reversing the previous question, how many different clusters does each star contribute to? 
Do all the different  arms of a star contribute dipoles to different clusters. If yes, 
it would definitely help form a percolating network of stars or a macroscopic gel of multiply-connected stars. 
Each star arm would be connected to many different stars through a dipole cluster.
Refer Figure \ref{fig5},\ref{fig5a} for discussions.
\item How would doubling the length of star arms help/hinder the cluster formation with dipoles ? A polymer 
arm could potentially explore more phase space and help forming bigger dipole clusters and help stars form networks.
\end{enumerate}

\begin{figure}[!htb]
\centering
\includegraphics[width=0.99\columnwidth,angle=0]{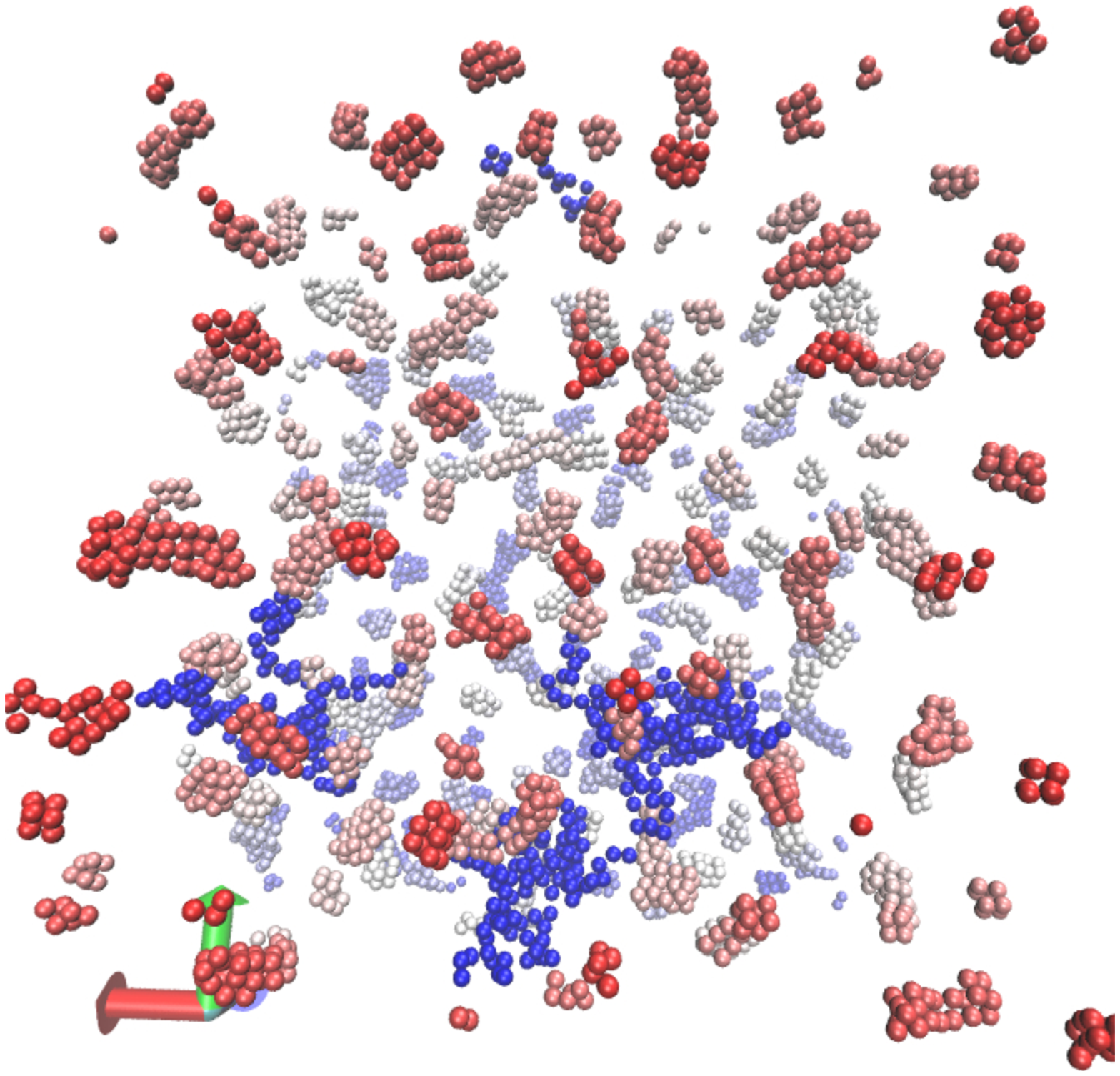} 
\includegraphics[width=0.99\columnwidth,angle=0]{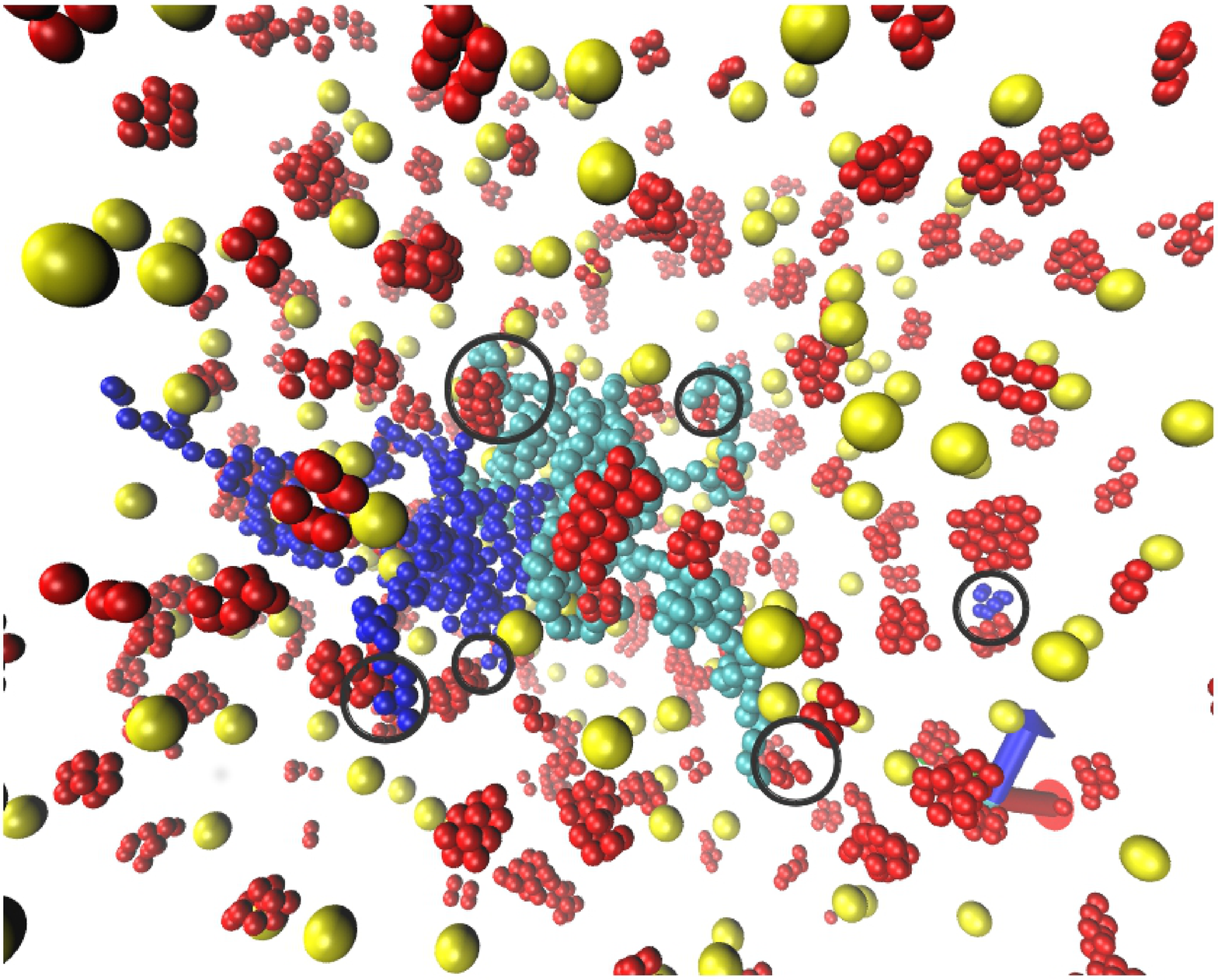} 
\caption{\label{fig0} Representative snapshots from our simulations of star polymers with $6$ arms 
and $f=6$ for $L=25$ monomers in each arm (top figure) and $L=50$ (bottom figure) at $R=40$.  
Each arm has a dipole at the tip of arms: $f=6$.  
We show uncharged monomers (small blue spheres) of only $3$ stars (top) and 
$2$ stars (bottom) out of the total stars present.  
All the charged monomers from each of the stars  are shown as slightly bigger red spheres with color
gradient in the direction pointing into the paper (top) but no color gradient in the bottom figure. 
Dipoles aggregate to form clusters, furthermore,  different star arms connect to different clusters. 
This is clearly seen in top panel, but we have marked by  circles in the bottom panel for the ease of reader.
The star centers, shown as yellow big spheres in the bottom panel, are homogeneously distributed over the 
simulation box.  In each panel, we see a few uncharged monomers of a star-arm isolated from the 
rest of star, this results from periodic boundary condition applied to the simulation box.  
}
\end{figure}

The primary quantities to vary are the effective charges $q e$ constituting the dipoles thereby changing 
interaction energy between dipoles and  the quantity $f$: the number of dipoles per star polymer. 
Instead of using fractional charge $qe$, we use quantity $R$ such that we can directly compare the
 thermal energy $k_BT$ and the electrostatic energy between $2$ charges at a distance of 
$\ell_0$ ($= 1 nm$) between them.  We have also considered 2 values of $L$, the number of monomers in 
each arm of the star polymer: we considered $L=25$ monomers per arm  as was used in experiments 
\cite{lele} as well as $L=50$ monomers.  
To compare results of $L =25$ and $L=50$ monomers per arm, we halve the number of stars to $175$ stars 
in the simulation box for the $L=50$ runs, thereby we  keep the number of monomers fixed. 
For reference and clarity, we give the relevant dipole/monomer numbers in Table. \ref{table1} 
as $L$ and $f$ is varied.  We keep the  number of arms per star remains fixed at $6$.
The arms can rotate freely about the center, hence, it is not relevant to discuss which particlular 
arm has the dipoles when $f=2$ and $f=3$.  For $f=6$, of course, each arm of each star has a dipole 
at the arm-tip.  In addition there are $350/175$ central beads of diameter $\ell_0^{sph}$ for 
$L=25/L=50$ star systems, respectively.

In Fig.\ref{fig0}, we show two snapshots from our simulations which show how star-arms contribute dipoles 
to different clusters. The top figure is for stars with $L=25$ monomers per arm; the bottom figure is for 
stars with $L=50$ with $f=6$. We have plotted the monomers (blue small spheres) of only $3$ and $2$
representative star-polymers, respectively, out of the 350 stars present in the box for ease of visualization.
All the dipoles from each of the arms of stars are shown in the snapshot to give the reader an  
idea of  spatial and size distribution of dipole clusters.  The dipoles are shown 
in red (slightly bigger spheres than monomers),  the red spherical aggregates indicate clusters 
of dipoles formed. One sees that the dipoles arrange to form elongated aggregates. This is presumably 
to arrange the dipoles anti-parallel to each other with negative charge next to the positive charge 
 and is in marked contrast to the chain like structures reported in \cite{dipole4,dipole5,dipole6} who 
use the dipolar approximation of the interaction potential. We have checked that 
for lower values of $R$, e.g. $R=20,10$ the dipoles cluster is shaped more like a globule rather than arranged 
in rod-like clusters. The snapshot at the bottom with $175$ stars has half the number of dipoles  than the
 snapshot at top. In bottom panel, we can see only part of the box as we have zoomed
in on the part which has stars for better clarity.  One can also visually analyze how the different arms 
contribute dipoles to dipole-clusters: in the bottom snapshot we have encircled 6 arms where different 
star-arms end up in different dipole clusters to help the reader.  The bottom snapshot also shows 
the distribution of star centers (big yellow spheres), and one can infer that they are relatively 
uniformly distributed in the system and there is no aggregation of star-centers around big clusters 
of dipoles. The qualitative conclusions arrived at from the snapshot is quantified in the figures that follow.

\begin{table}
\begin{tabular}{ |p{1cm}|p{2.75cm}|p{1cm}|p{1cm}| } 
 \hline
L & $S_T=$ No. of Stars & f & $N_D$ \\
\hline
\hline
25 & 350 & 6 & 2100  \\ \hline
25 & 350 & 3 & 1050  \\ \hline
25 & 350 & 2 & 700  \\ \hline
50 & 175 & 6 & 1050  \\ \hline
50 & 175 & 3 & 525  \\ \hline
50 & 175 & 2 & 350  \\ \hline
\end{tabular}
\caption{ Table listing the number of dipoles in the simulation box for as $L$ and $f$ is varied. The 
total number of monomers always remains fixed at $L*f*S_T = 52500$. There are $175/350$ central monomers 
if there are $175/350$ stars in simulation box. Number of dipoles $N_D = f*S_T$.}
\label{table1}
\end{table}

Figure \ref{total} quantifies the ideas presented in Fig.\ref{fig0} and shows the average of total 
number of dipole-clusters  $C_T$ as a function of $R$ for different values of $f$.
We define $2$ dipoles to belong to the same  cluster if the distance between the center of $2$
charged monomers belonging to different stars arms is less than $1.2 \sigma$.  
A cluster of size $1$ indicates there is only a dipole in the cluster implying that the dipole 
has not formed an aggregate with another dipole. 
Data presented in Fig. \ref{total} is for $L=25$ and $L=50$ monomers per arm in subplot 
(a) and (b), respectively. For $R=5$ and $R=10$, the total number of clusters $C_T$ for all $f$s
is only slightly less than the total number of dipoles $N_D$ in each case (refer Table\ref{table1})
implying that most dipoles are isolated and free in space. Most of the cluster are essentially 
a $1-$dipole cluster, and very few clusters have $2$ dipoles. 

In contrast, for $R=20$ and $R=40$, the average number of clusters $C_T$ in the box are much smaller
than $N_D$, indicating that majority of dipoles are in clusters and each cluster has multiple dipoles.
Refer Table \ref{table1} for the values of $N_D$.  The value of  $C_T$ calculated using the data 
of Fig.\ref{fig1} and \ref{fig1a}: $C_T = \sum C \times n_c$, where the summation is over 
the number of dipoles $C$ in a cluster.  The average number of clusters  containing $C$ dipoles 
is denoted by $n_c$: $n_c$ versus $C$ data is discussed in the next paragraph.  
Note that the number of clusters keeps fluctuating as dipoles aggregate to form clusters and 
then break apart due to thermal energy. In addition, when the system is heated to $50 k_BT$, 
every cluster disintegrates completely, and a new distribution of dipole-clusters 
is formed once the system is cooled down. Thus $n_c$, and thereby $C_T$ is a 
statistically averaged quantity. In the data presented in subplot (b) for $L=50$,
we again observe that $C_T/N_D \approx 1$ for $R=5,10$, whereas as one increases $R=20,40$ 
there is aggregation of dipoles to form larger clusters.

\begin{figure}[!htb]
\centering
\includegraphics[width=0.49\columnwidth,angle=0]{./Total_clusters/L25_Totalclusters.eps} 
\includegraphics[width=0.49\columnwidth,angle=0]{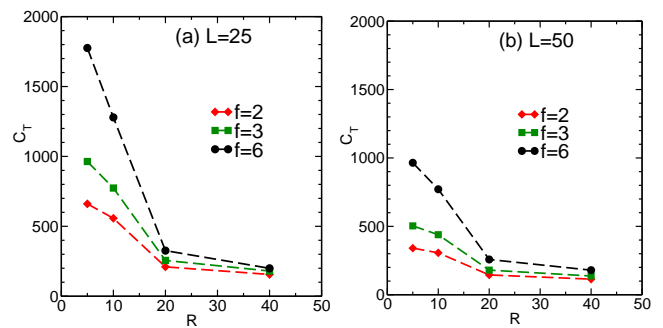} 
\caption{\label{total} Figure shows the average of total number of clusters $C_T$ in the system versus $R$,
 for different values of $f$, the number of dipoles per star. Subplots (a) and (b) 
are for $L=25$ and $L=50$ with $350$ and $175$ stars in the simulation box, respectively. 
 The quantity $R$ is the ratio of energy $E(qe, 1nm)$ and $k_BT$, where $E(qe, 1nm)$ is the energy 
between two partial charges $\pm qe$ at a distance 1nm from each other. 
There are two distinct regimes:
($i$)for $R=5$ or $R=10$ the value of $C_T \sim N_D$, where $N_D$ is the total number of dipoles in the system.
This indicates that most clusters have just $1$ dipoles. ($ii$) For $R=20$ or $R=40$, $C_T \ll N_D$,
 indicating the each cluster has a large number of dipoles aggregated together in a  cluster.  
Data presented is over $4$ rounds of heating and cooling cycles, and after each cycle 
a statistically independent set of clusters are formed. 
Data has not been normalized by $N_D$, since this data will be used later for  analysis.
}
\end{figure}

Figure\ref{fig1} shows the distribution of the number of dipole-clusters $n_c$ for a particular  
size of the cluster $C$; the size $C$ of dipole-cluster is calculated by the number of dipoles in
 the particular cluster.  For $R=40$ and $f=6$, all the dipoles at the end of star-arms 
 have aggregated to form clusters: there are no clusters of size $1$: refer subplot \ref{fig1}(a). 
In Fig.\ref{fig1}a, there are on an average $5$ large clusters containing $C=15$ dipoles in each cluster. 
Clusters containing $16/17/18$ dipoles occur in the box with similar frequency.  Even bigger clusters 
with more than $20$ dipoles per cluster are seen though with less frequency for
$R=40$.  Clusters  with $6-10$ dipoles per clusters are found with the highest frequency as shown in the peak 
of the distribution.  For lower values of $f=3$ and $f=2$ dipoles per star, the dipole density is 
lower in the box, and the peak of the distribution shifts to lower values of cluster size. In these cases, 
one sees large  number of clusters with just $4$ or $5$ dipoles in a cluster. It is difficult to confirm if 
larger aggregates will form over much longer time scale of simulations, but we expect our present result to
hold true. This is  because diffusion of individual stars will be hindered because different arms
of stars are in different clusters (as we show later) with relatively high values of Coulomb 
energy $C_E/k_BT$ per dipole. Furthermore, large aggregates of dipoles with 
$25$ (say) dipoles or more per cluster for $f=2$,$f=3$ will also result in local increase in 
density of the stars connected to the cluster, i.e, spatially inhomogeneous monomer density.

\begin{figure}[!htb]
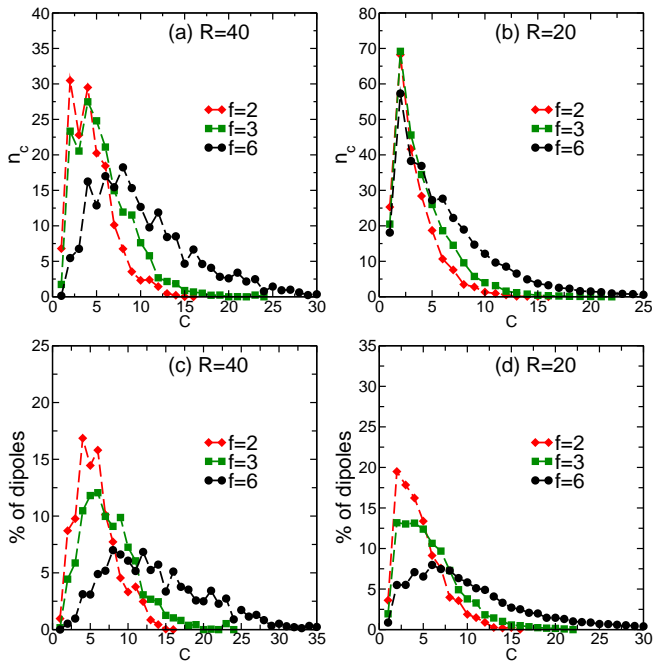

\centering
\includegraphics[width=0.49\columnwidth,angle=0]{./Clusters_vs_dipoles/L25_S40_Varyf.eps}
\includegraphics[width=0.49\columnwidth,angle=0]{./Clusters_vs_dipoles/L25_S20_Varyf.eps}\\
\vskip0.1cm
\includegraphics[width=0.49\columnwidth,angle=0]{./Clusters_vs_dipoles/L25_S40_Varyf_norm.eps}
\includegraphics[width=0.49\columnwidth,angle=0]{./Clusters_vs_dipoles/L25_S20_Varyf_norm.eps}
\caption{\label{fig1} 
The plots (a),(b) shows the distribution of the average number of  clusters $n_c$ of  cluster size $C$ 
for $R=40$ and $R=20$.  The quantity $C$ is the number of dipoles in a cluster.
The number of dipoles at tip of arms in a $6$ arm star polymer is denoted by symbol $f$, and we show 
data for different values of $f$ for the number of monomers per star-arm $L=25$. There are $350$ stars in
the simulation box. Subplots (c) and (d) show the percentage of dipoles which are to be found in clusters 
of size $C$ for the corresponding set of parameters of (a) and (b) respectively. 
We have not suitably normalized the $y$ axis for reasons given in the text.
}
\end{figure}

In figure \ref{fig1}(b), for $R=20$ with $f=6$, one observes that relatively smaller 
clusters are formed compared 
to when $R=40$ and the peak of distribution has shifted to lower values of $C$. 
 One concludes that there is lesser effective attraction between the dipoles to form 
large clusters and thermal energy destabilizes clusters with $20$ or more dipoles per cluster.  The
 peak of the distribution for $f=6,3,2$ all lie at around $2$ dipoles per cluster. However, for $f=6$, 
there are  more than $10$ clusters each of size $10,11,12$, i.e. $n_C>10$ for $C=10,11,12$ each; so 
actually a large fraction of the dipoles can be expected to reside in  clusters of size $\geq 10$. 
The fraction  of total dipoles which are in clusters of size $C$ are plotted in Fig. \ref{fig1}c 
and \ref{fig1}d for $R=40$ and $R=20$.  There are also a few large clusters (with $C>10$) for $R=20$, so 
actually around $45 \%$ of the dipoles are in clusters with $C\geq 10$ for $R=20$ with $f=6$. 
For $f=2$ and $f=3$ cases, the number of clusters with $C \geq 10$ decreases along with the 
fraction of dipoles in such clusters. This is obviously due to fewer number of dipoles present in 
the system for $f=2,f=3$. But still, independent of the value of $f$, 
at least $50 \%$ of the dipoles are in clusters with $6$ or more dipoles for $R=40,20$.
This also implies that $50\%$ of the total number of stars arms contribute dipoles to clusters of size 
$C\ge6$. 

\begin{figure}[]
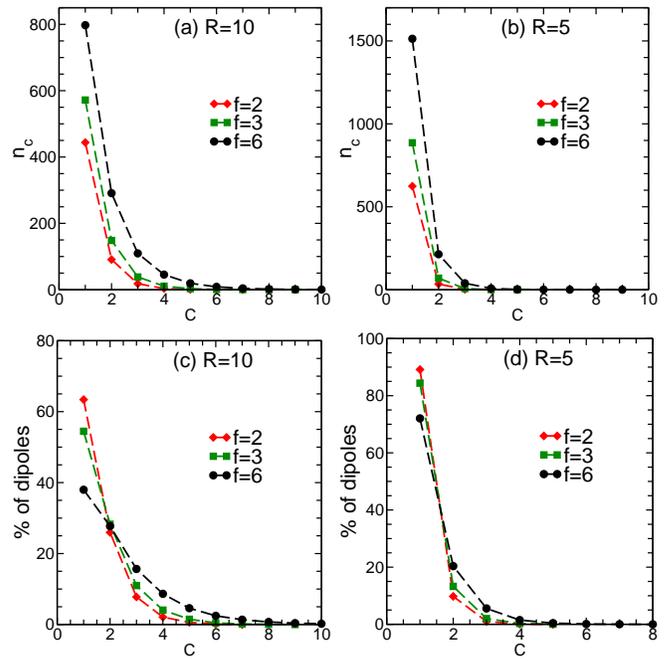

\centering
\includegraphics[width=0.49\columnwidth,angle=0]{./Clusters_vs_dipoles/L25_S10_Varyf.eps} 
\includegraphics[width=0.49\columnwidth,angle=0]{./Clusters_vs_dipoles/L25_S5_Varyf.eps} \\
\vskip0.1cm
\includegraphics[width=0.49\columnwidth,angle=0]{./Clusters_vs_dipoles/L25_S10_Varyf_norm.eps} 
\includegraphics[width=0.49\columnwidth,angle=0]{./Clusters_vs_dipoles/L25_S5_Varyf_norm.eps} 
\caption{\label{fig1a} 
The plots (a),(b) shows the distribution of the average number of clusters $n_c$ versus the cluster size $C$
for lower values of $R$, viz., $R=10$ and $R=5$ with different values of $f$. In contrast to Fig. \ref{fig1},
dipoles with $R=10$ or $R=5$ form very small aggregates with just one or two dipoles per cluster, i.e. the distribution
is sharply peaked at $C=1$ and $C=2$.  There are $350$ stars in the box with $L=25$ monomer per arm.
Subplots (c) and (d) show the percentage  of total number of dipoles which are observed in clusters of size $C$.
We have not suitably normalized the $y$ axis for reasons given in the text.
}
\end{figure}

In Figures \ref{fig1}a and $b$, we do not normalize the $y$-axis by either the total number of dipoles 
$N_D$ or by the average of total number of clusters $C_T$. The reason is that the total number of 
clusters is not a conserved quantity and does change due the course of the run. The total number of
 dipoles does remain fixed at $N_D=2100,1050$ or $700$ for $f=6,3$ or $f=2$, respectively, but then 
if the y-axis get divided by $N_D$, one looses the estimate of the number of clusters, 
especially for $f=6$ where $n_c << N_D$.

\begin{figure*}[]
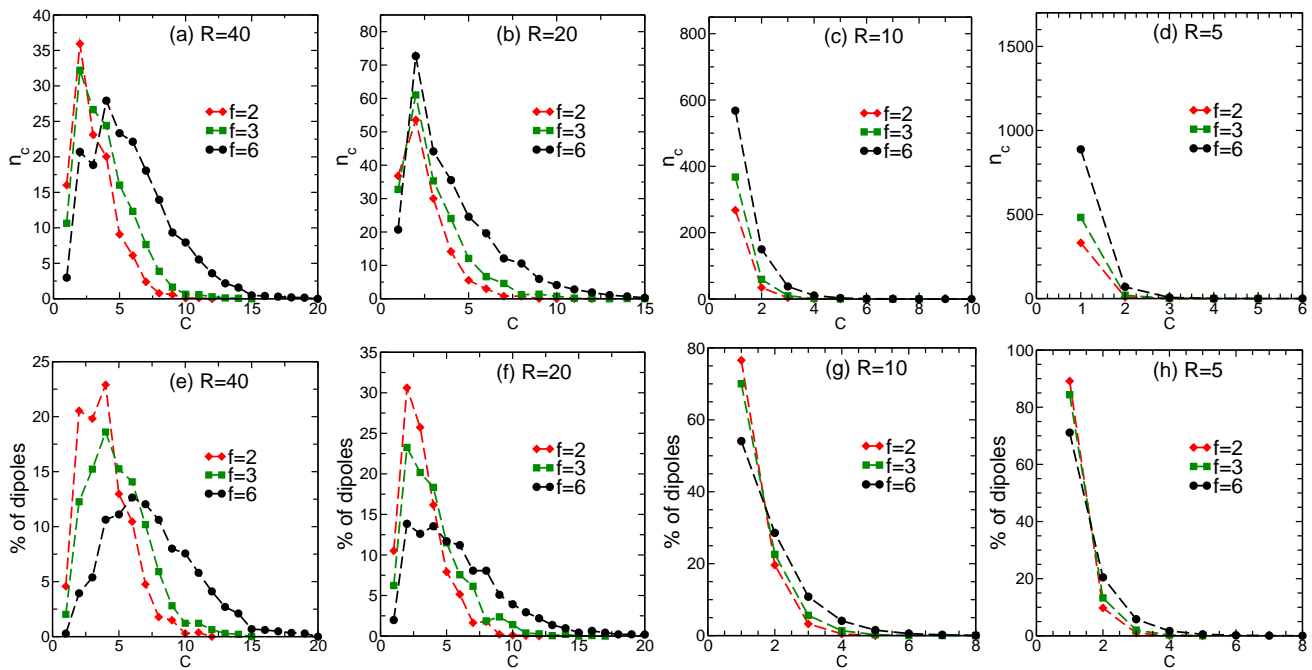

\centering
\includegraphics[width=0.49\columnwidth,angle=0]{./Clusters_vs_dipoles/L50_S40_Varyf.eps}
\includegraphics[width=0.49\columnwidth,angle=0]{./Clusters_vs_dipoles/L50_S20_Varyf.eps}
\includegraphics[width=0.49\columnwidth,angle=0]{./Clusters_vs_dipoles/L50_S10_Varyf.eps} 
\includegraphics[width=0.49\columnwidth,angle=0]{./Clusters_vs_dipoles/L50_S5_Varyf.eps} \\
\vskip0.1cm
\includegraphics[width=0.49\columnwidth,angle=0]{./Clusters_vs_dipoles/L50_S40_Varyf_norm.eps}
\includegraphics[width=0.49\columnwidth,angle=0]{./Clusters_vs_dipoles/L50_S20_Varyf_norm.eps}
\includegraphics[width=0.49\columnwidth,angle=0]{./Clusters_vs_dipoles/L50_S10_Varyf_norm.eps} 
\includegraphics[width=0.49\columnwidth,angle=0]{./Clusters_vs_dipoles/L50_S5_Varyf_norm.eps} 
\caption{\label{fig2} 
The plots (a),(b),(c) and (d) shows the distribution of the average number of  clusters $n_c$ 
versus the  cluster size $C$, for stars with $L=50$ and $175$ stars in the simulation box for 
$R=40,20,10,5$, respectively. The quantity $n_C$ is the average number of clusters in the simulation
 box with $C$ dipoles in a cluster. The figures at bottom (e),(f),(g),(h) 
show the percentage of total number of dipoles which are found in clusters of size $C$, so that we can estimate
what fraction of dipoles are found in big/small clusters.
}
\end{figure*}

The cluster size distribution is very different when $R=10$ and $R=5$, refer (a) and (b) of Fig. \ref{fig1a}.
The attraction between dipoles is hardly enough to bring dipoles together to form big aggregates, 
i.e., a cluster with more than 6 dipoles ($C>6$, say) in a cluster.
Most dipoles are free in space, a small number form dimers due to Coulomb attraction: 
thus entropy wins over Coulomb attraction between dipoles.
For $R=10$, there are  more than $70\%$ of dipoles in clusters which contain a single dipole or two dipoles
($C=1$ or $C=2$) for $f=6$. 
The percentage of dipoles in a cluster with $C=1$ increases for $f=3$ and $f=2$: refer 
Fig.\ref{fig1a}(c). In Fig.\ref{fig1a}d, we see that for $R=5$ and $f=6$, less than $10\%$ of all dipoles form 
dipole-trimers or bigger clusters. 
This percentage reduces to nearly zero for $C=3$ (or more) dipoles per cluster when $f=2$ or $ f=3$.


These observations leads to the question that if the star polymers have longer arms, 
would that help forming bigger clusters of dipoles especially for $f=2, f=3$. 
One can imagine that different star-centers could be spread out in space 
and yet with the advantage of longer star arms and hence more reach, the dipoles could aggregate to 
form bigger clusters than when $L=25$.  As mentioned before, we do simulations with $175$ stars
 with $6$ arms each, but with $50$ monomers per arm, i.e. $L=50$ and compare it with
the $L=25$ data. The distribution of cluster sizes  $n_C$ versus $C$
is presented in Fig.\ref{fig2}  for different values of $R$ and $f$. 
The number of monomers remain the same but the number of dipoles
 at arm tips is halved, such that $N_D=1050$. So data for simulations
 of $L=50,f=6$ system could be compared with $L=25,f=3$ data as they have the same  
dipole-density.  

On comparison of Fig.\ref{fig2} with  Fig.\ref{fig1} and Fig.\ref{fig1a}  we observe that the data for size 
distribution of clusters, i.e. $n_C$ versus $C$ are nearly the same for 2 systems with 
 $2$ different length of arms, but  with identical $N_D$!
For example, comparison of $n_C$ vs. $C$ of stars of length $L=25$ with $f=3$ and 
stars of length $L=50$ with $f=6$ shows that they are nearly identical, that too for each value of $R$. 
It implies that the doubling the length of polymer arms has no effect on the cluster size-distribution 
of dipoles, and cluster size distribution is decided primarily by the $R$ value and the number density of 
dipoles in the box.  Similar consistent behaviour can be observed if we compare cases $f=2, L=25$ 
and $f=3, L=50$ though the number density of dipoles is not exactly the same in these two cases. The fraction
of dipoles to be found in clusters of size $C$ are shown in figure \ref{fig2}e,f,g,h for $R=40,20,10$ 
and $5$, respectively.

\begin{figure}[htb]
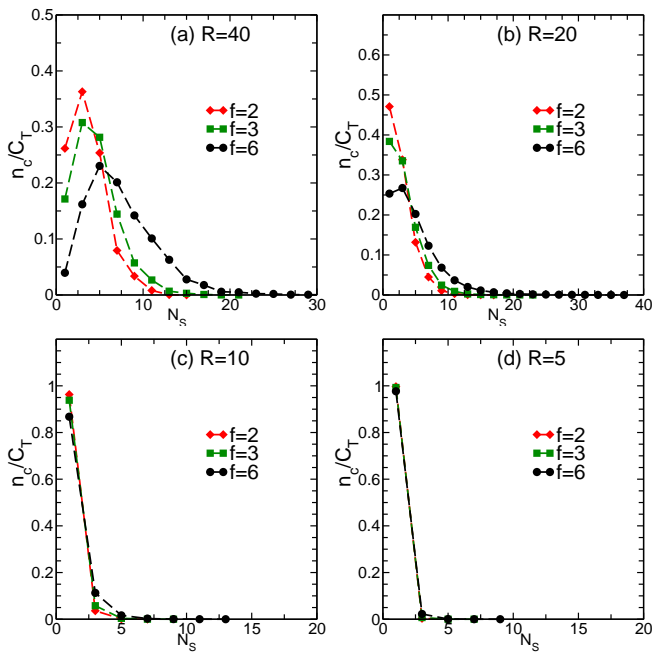

\centering
\includegraphics[width=0.49\columnwidth,angle=0]{./Clusters_vs_Starbin/L25_S40_Varyf_norm.eps} 
\includegraphics[width=0.49\columnwidth,angle=0]{./Clusters_vs_Starbin/L25_S20_Varyf_norm.eps}  \\
\vskip0.1cm
\includegraphics[width=0.49\columnwidth,angle=0]{./Clusters_vs_Starbin/L25_S10_Varyf_norm.eps} 
\includegraphics[width=0.49\columnwidth,angle=0]{./Clusters_vs_Starbin/L25_S5_Varyf_norm.eps} 
\caption{\label{fig3} 
The plots show the distribution of the number of clusters $n_c$, normalised by the average of the 
total number of clusters $C_T$,  versus $N_{S}$, the number of stars which contribute dipoles 
to make up a cluster. 
Data is for $L=25$ with $350$ stars in simulation box for different values of $f$, the number of arms in a star
with dipoles at arm-tips.  For $R=5,R=10$, the dipoles do not form clusters with multiple dipoles 
in cluster and hence has contributions primarily from just one star. In contrast for $R=40$, each 
cluster has dipoles from multiple stars. Note that the bin size in the $x-$axis is $2$. 
}
\end{figure}
In general, we make the following observations about the $L=50$ system:\\
$\bullet$ For $R=5$, $90\%$ of dipoles with  $f=2$ stars and nearly $80 \%$ of dipoles  of stars with $f=6$ 
are isolated single dipoles. This is seen from the data at $C=1$ in Fig.\ref{fig2}d,h.\\
$\bullet$ As $R$ value is increased from $R=5$, the dipolar attraction between dipoles increases to 
gradually overcome the thermal effects.  At $R=20$ with $f=6$,  more than $50 \%$ of the dipoles are 
in clusters of size $C\ge5$: refer Fig.\ref{fig2}f.  The peak of the size distribution data in 
Fig.\ref{fig2}b is at $C=2$, but still more than a third  of the total number of clusters 
($C_T \approx 240$, refer Fig.\ref{total}b) are clusters with $C\ge5$.  Thus one could expect a large 
fraction of star-arms to be inter-linked by dipole clusters and the system could be in a gel state, 
if a cluster gets dipole contributions from different stars. \\
$\bullet$ For $R=20$ with $f=3$, the average number of clusters in the system is $C_T \approx 200$ 
(refer Fig.\ref{total}b), and only one-fifth of dipoles (and only $1/8$ of the clusters) have the 
number of dipoles in cluster of size $C \ge 5$.  \\
$\bullet$ For $R=40$ with stars with $f=6$, nearly $65 \%$ of dipoles are in clusters with size 
$C \ge 5$, moreover, $50 \%$ of the total number of clusters have $5$ or more dipoles. 
These fractions obviously are lower when $f=2$ or $f=3$. 

\begin{figure}[!htb]
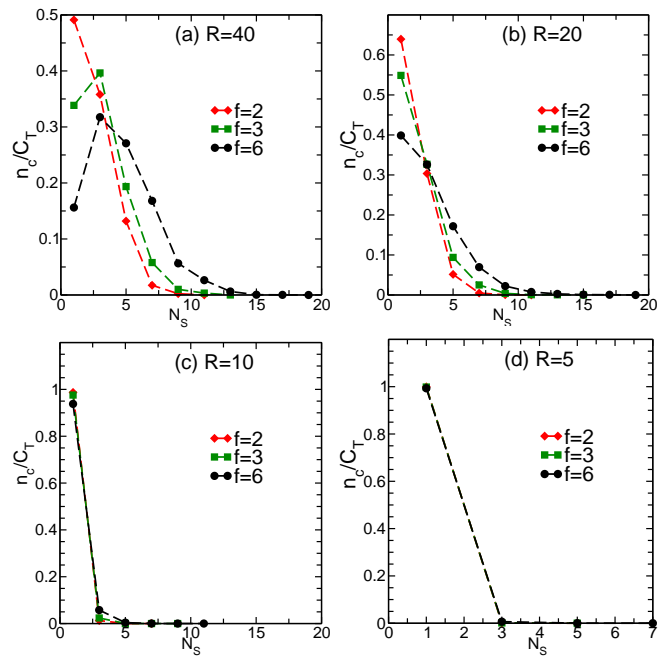

\centering
\includegraphics[width=0.49\columnwidth,angle=0]{./Clusters_vs_Starbin/L50_S40_Varyf_norm.eps} 
\includegraphics[width=0.49\columnwidth,angle=0]{./Clusters_vs_Starbin/L50_S20_Varyf_norm.eps}  \\
\vskip0.1cm
\includegraphics[width=0.49\columnwidth,angle=0]{./Clusters_vs_Starbin/L50_S10_Varyf_norm.eps}
\includegraphics[width=0.49\columnwidth,angle=0]{./Clusters_vs_Starbin/L50_S5_Varyf_norm.eps} 
\caption{\label{fig4} 
The plots show, the distribution of the number of clusters $n_c$, normalised by the average of 
total number of clusters $C_T$, are plotted versus $N_{S}$, the number of stars which contribute dipoles 
to make up a cluster.  Data is for $L=50$ with $175$ stars in simulation box for different values of 
$f$, the number of arms in a star with dipoles at arm-ends. 
For $R=10,R=5$, the dipoles do not form clusters with multiple dipoles, thus $n_C/C_T \approx 1$
for $N_S =1$. In contrast for $R=40$, each cluster has dipole contributions from multiple stars.
}
\end{figure}

The next question to investigate is that, given that one has quite a few clusters with $C\ge5$ and large 
fraction of the total number of dipoles in these clusters, how many stars is each cluster connected to?
These questions is systematically investigated in fig \ref{fig3} and \ref{fig4}, where the distribution of the 
number of clusters $n_c$, normalized by the corresponding total number of cluster $C_T$,
 is plotted versus the number of stars $N_S$ that the dipole-clusters 
gets dipole contributions from.  Figure \ref{fig3} and \ref{fig4} is for $L=25$ and $L=50$, 
respectively. 

For $R=40, f=6$ and $L=25$ stars, which has $C_T \approx 250$ (refer Fig. \ref{total}a), 
three-fourth  of the total number of clusters $C_T$ have  $5$ or more dipoles 
in a cluster (refer Fig.\ref{fig1}a).  Now we can add the values of $n_c/C_T$ 
for different values of $N_S$  in Fig.\ref{fig3}a to see that a predominantly large number
 of clusters, more than $70 \%$,
 have contributions from $5$ or more stars: the peak of the $n_C/C_T$ distribution is at $N_S = 5,7$. 
Moreover, more than $20\%$ of the clusters have $N_S>10$, i.e., around $50$ out of $250$ (approximately) 
clusters have dipole contributions from more than $10$ stars!  For $f=2$ and $f=3$ stars 
large dipole-clusters are unable to form and hence  clusters gets contributions from  fewer 
number of stars $N_S$ compared to the $f=6$ system. But still $40\%$ of the clusters
have dipole contributions from $5$ or more  stars for $f=2,f=3$.  

\begin{figure}[!htb]
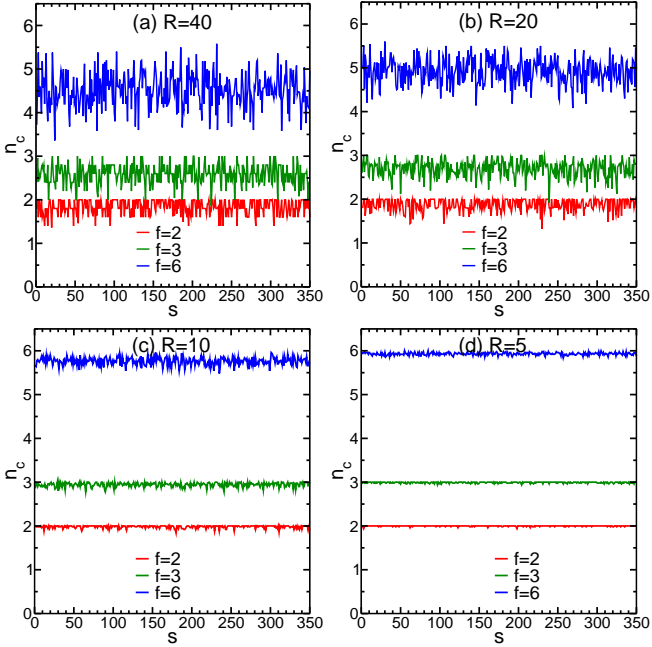

\centering
\includegraphics[width=0.49\columnwidth,angle=0]{./Clusters_vs_Stars/L25_S40_Varyf.eps} 
\includegraphics[width=0.49\columnwidth,angle=0]{./Clusters_vs_Stars/L25_S20_Varyf.eps} \\
\vskip0.1cm
\includegraphics[width=0.49\columnwidth,angle=0]{./Clusters_vs_Stars/L25_S10_Varyf.eps} 
\includegraphics[width=0.49\columnwidth,angle=0]{./Clusters_vs_Stars/L25_S5_Varyf.eps} 
\caption{\label{fig5} 
The subplots show the average number of clusters $n_{c}(s)$ that each star contributes its dipoles to.
The x-axis shows the star index $s$, the simulation box has 350 stars, and the stars are numbered (indexed) 
$1$ to $350$.  Subplots (a),(b),(c),(d) correspond to  the values of $R=40,20,10,5$ respectively for 
case $L=25$ monomers in a arm.  
 For $R=5$ and $f=6$, each dipole is isolated and therefore is in a 
different cluster with just 1 dipole in cluster. Each star contributes to $6$ clusters.
However, when $R=40$  we know that each cluster contains multiple dipoles, thus for $f=6$ the data shows 
that $6$ different arms contribute  dipoles mostly to different clusters. 
This definitely helps in forming a gel-like network of polymer arms across the box. 
}
\end{figure}
For $R=20$ with $f=6$ stars (Fig.\ref{fig3}b), where one obtains larger number of  smaller clusters,
but nearly half of the total number of clusters are connected to more than $5$ stars with $N_s \ge5$ 
and nearly $75 \%$ of the clusters are connected to $3$ or more stars. 
Thus each dipole cluster acts as a node through which different stars 
are connected by contributing a dipole from one (or more than one) star arm. Other arms of the star could be connected 
to a different cluster (we systematically investigate this later in the text). Thus there is a possibility
of forming a system spanning percolating network of a polymer gel. 
Even if a  percolating network of polymers is not formed for $R=20$,  the star polymers would effectively 
form very large macromolecules connected through the dipole clusters.  
There exists few  clusters with upto $14$ or $16$ dipoles per cluster
which get contributions from upto $N_S=14$ stars thus form a very large macromolecule as each star 
in turn will be connected to other dipole clusters.
This can be independently deduced and is consistent with  data of Figs. \ref{total}a and \ref{fig1}c.

For $R=20$ with $f=2,3$ dipoles per star, one might not get percolating gels or large macromolecules held together
by dipole clusters, but stars do get conjoined and form polymers of {\em effective} larger molecular weight
than individual stars.  More than half the clusters have contributions from $3$ or more stars.
For lower values of $R$, viz, 
$R=10$ and $R=5$ (refer Fig.\ref{fig1a}a,b), most of the clusters have $1$ or $2$ dipoles, thereby each cluster 
has dipole-contributions from $1$ or $2$ stars: the stars are hardly networked with each other through
 dipole clusters. Thereby $R=10, R=5$ stars do not form a gel-like network of polymers, and this is seen 
in Fig\ref{fig3}c and \ref{fig3}d where more than $90\%$ of clusters are connected to $1$ or $2$ stars: there 
are sharp peaks for $N_s =1$.  This result is independent of the value of $f$, the Coulomb interaction between 
dipoles $R$ is not enough such that clusters containing many dipoles get aggregated into larger macromole.
For $R=10, f=6$, there are $10\%$ clusters which are connected to $2$ stars, that is effectively doubling the 
{\em effective} molecular weight of these stars which could lead to an effective marginal increase in viscosity.
 
\begin{figure}[!htb]
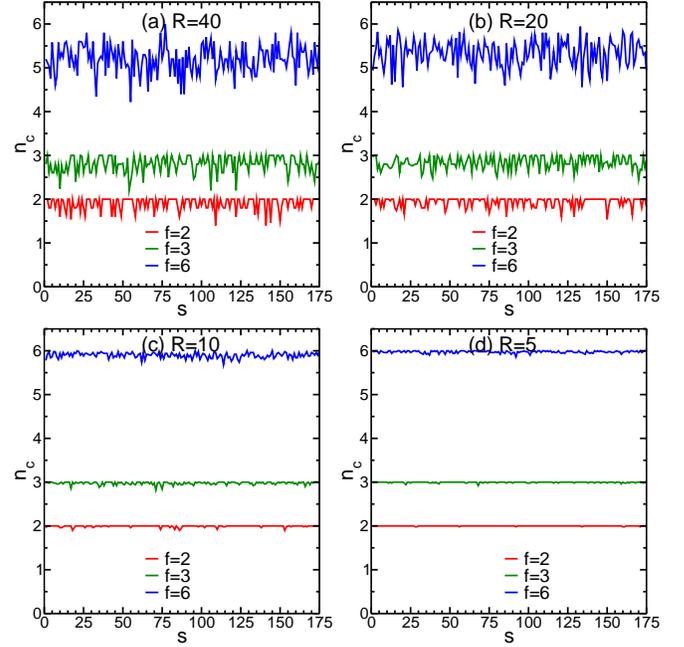

\centering
\includegraphics[width=0.49\columnwidth,angle=0]{./Clusters_vs_Stars/L50_S40_Varyf.eps} 
\includegraphics[width=0.49\columnwidth,angle=0]{./Clusters_vs_Stars/L50_S20_Varyf.eps} \\
\vskip0.1cm
\includegraphics[width=0.49\columnwidth,angle=0]{./Clusters_vs_Stars/L50_S10_Varyf.eps} 
\includegraphics[width=0.49\columnwidth,angle=0]{./Clusters_vs_Stars/L50_S5_Varyf.eps} 
\caption{\label{fig5a} 
The subplots show the average number of clusters $n_{c}(s)$ that each star contributes its dipoles to.
The x-axis shows the star index $s$, the simulation box has $175$ stars, and the stars are numbered (indexed) 
$1$ to $175$.  Subplots (a),(b),(c),(d) correspond to  the values of $R=40,20,10,5$ respectively for 
case $L=50$ monomers in a arm. 
}
\end{figure}

For $L=50$ with fewer dipoles in the box, the results and conclusions for $R=5$ and $R=10$ stars are not 
any different from the $L=25$ stars with $R=5,10$ as seen in Fig.\ref{fig4}c and d.
For $R=40$ stars the star-polymers get networked  when $f=6$ and $f=3$  as nearly 
$50 \%$ and $35 \%$ of the clusters have dipole contributions from more than $4$ stars, respectively. 
For $f=6$, one could probably get a percolating gel of stars, but $f=2$ or $f=3$, 
though the stars do get {\em gelled} through dipole clusters, it is unlikely that the 
gel will be system spanning/percolating through the system. But one would definitely observe a
large increase in effective relaxation-times/viscosity due to the slow dynamics of effectively 
large macromolecules formed. 

For $R=20, L=50$, refer Fig.\ref{fig4}b, there is very little chance of stars with $f=2$ or $f=3$ dipoles
to form percolating gel structure of stars. 
For $f=6$, the system could form a percolating gel as nearly $30\%$ of the clusters are connected 
to more than $5$ stars, thus $30\%$ of clusters have $5$ or more dipoles. 
 This can be independently checked from Fig.\ref{fig1a}b, moreover, $50\%$ of dipoles are clusters with 
$C\ge5$ (ref Fig.\ref{fig1a}f). 
At the very least, multiple stars get connected through dipole clusters and form large aggregates of stars.  
This should lead to significant increase in viscosity through we are do 
not try to quantify the value of viscosity with the computational resources presently available to us. 
 The other interesting thing to 
note is that the data for $L=50,f=6$ is quantitatively very similar to to the case of $L=25,f=3$ 
for all values of $R$. As mentioned before the two systems have identical number of dipoles.
As we have seen before, there doubling the length of star arms from $L=25$ to $L=50$ seems to have no effect on the 
size distribution of dipoles clusters or the the way clusters get dipole contributions  from different stars.

Next we aim to calculate and find out  how many clusters does each star contribute arms/dipoles to? 
Figures \ref{fig5} and \ref{fig5a} show the average number of clusters $n_c$ (on the y-axis) that 
each star contributes dipoles to. Data for Figs.\ref{fig5} and \ref{fig5a} is for $L=25$ and $L=50$, respectively.  
A particular star $s$ could contribute to $3$, $4$ or $5$ clusters at 
different times of the simulation run, thereby we get non-integer values of $n_c$. 
For $R=40,f=6$ (Fig. \ref{fig5}a) for $L=25$, each star is connected on an average to between $4$ and $5$ dipole clusters. 
This  now confirms our previous understanding that this system corresponds to a system spanning star gel-like network. 
As mentioned before and shown in figures \ref{fig3}, $70 \%$ of clusters have contributions from more than
$5$ stars. Hence, each star would be contributing multiple arms to such {\em multiply-connected dipole clusters},
and this would lead to a system-spanning gel like molecular network. 
With each star connected to clusters with many dipoles, the stars will form a system 
spanning network of star arms: corresponding to a physical gel. Obviously the energy/stress needed to shear such a 
polymer gel would be very large compared to a system of {\em unconnected} star polymers. This will correspond to 
a large increase in $G^{'}$ and $\eta$ as seen in rheology experiments of stars with ionomers at star tips.

For lower values of $f$ with $R=40$, which has fewer large clusters, each arms typically is 
connected to different dipole-clusters.
But it does  happen, though rarely, that both the arms with the dipoles of a $f=2$ star ends 
up in the same dipole cluster: the average $n_c$ does show values less than $2$ for $R=40$. For almost all 
values of $f$,$R$ and $L$ , the value of $n_c(s)$ is close to the value of $f$ for different set of parameters
for almost all stars. This indicate that 
the different arms of stars mostly go to different dipole-clusters and unlikely to aggregate together. 
For $f=6$ and $L=25$ stars, $2$ out of $6$ different arms do end up in the same cluster, the values 
of $n_c$ in Fig.\ref{fig5}a are between $4$ and $5$. 

However, using our 
understanding of the previous figures, we must interpret data for $R=5, R=10$ very differently from $R=40,R=20$.
For $R=10,5$, we know that there are no large clusters and each dipole is nearly free and isolated, so of course
each star is shown to contribute each arm to different clusters. On the other hand for $R=40, R=20$, we know 
that a very large proportion of clusters are clusters with $5$ or more dipoles, so most of the star arms connect 
to such multi-dipole clusters. As mentioned before, this indicates that each star will be multiply connected with 
different stars through an average of $4$ different dipole clusters, and sometimes $2$ arms from the same star 
can land up in the same dipole cluster. A few arms of stars will of course be free and not connected to clusters.  
Note that for $L=50$ with $R=40,R=20$ and $f=6$, the values of $n_c$ is larger than that for the corresponding 
$L=25$ studies. Our understanding is that the $L=50$ stars have smaller number of clusters, and it is more possible  
for a star arm to end up as a free arm with cluster size $C=1$.

\section{Discussion}

To conclude, we establish by molecular dynamics simulation of a melt of star polymers with dipoles at the tip 
of star-arms that for values of effective charge $qe=0.57e$ and $qe=0.81e$ ($R=20$ and $R=40$, respectively
at $T=300K$) of effective charges we get a high degree of network formation through 
the formation of dipole clusters. Typically, each arm  of a star is connected to a different dipole cluster. 
Each dipole cluster for $f=6$ has contributions from many stars so the system should be nearly a percolating gel.
A small fraction of arms may remain free for stars with just $2$ or $3$ dipoles per star. 
Given that the energy per dipole is quite high compared to $k_BT$, it would be difficult to 
break the clusters. Data suggests that it is likely to be a percolating gel for $R=40$, but could
be a non-percolating gel for $R=20$ especially for lower values of $f$. For $f=2,f=3$,
the network of stars will result in effective macomolecules of large molecular weight. 
Such network formation of stars through dipole clusters should lead to a large increase in 
viscosity when the star-polymer melt is sheared. The network formation or physical gelation 
resulting in an  increase in viscosity of the melt is more for $f=6$ stars compared to 
$f=2,f=3$ stars, in tune with experimental observations. 

For lower values of $R$, i.e. $R=10$ and $R=5$, the dipoles at star arms do not aggregate 
sufficiently to form a physical gel. However, there is some clustering resulting in effective
increase of molecular weight of networked stars for $R=10$.  The average energy per dipole
is $5k_BT$ at $k_BT=1$ for $R=5$, that is same as  the energy of an isolated dipole showing 
the effective attractive interaction between dipoles is negligible.  If the diameter 
of monomers is smaller than $1 nm$, one could get clusters at significantly lower 
values of $qe$. This is because the interaction energy between 2 interacting dipoles should 
increase as oppositely charged monomers can approach each other to smaller distances.

The other interesting observation is that doubling the length of star arms has no effect in the 
distribution of size of dipole clusters, if we keep the uncharged monomer density as well as the
dipole density the fixed. So it is more useful to have telechelic stars with relatively shorter arms
with  large $f$, if one wants to increase effective viscosity of a polymer through this mechanism.

Finally, because we have modelled dipoles by explicit charges instead of effective dipolar interaction, 
the arrangement of dipoles in the dipole clusters of our simulations is very different from 
previously known studies \cite{dipole4,dipole5,dipole6}.  
For $R=40$, dipoles arrange themselves anti-parallel to each other 
in a row and form elongated aggregates as that is the low energy configuration compared to the 
dipoles lining up with every dipole moment approximately
pointing in the same direction \cite{dipole4,dipole5,dipole6}. It could be relevant to revisit 
current understandings regarding structure formation in dipolar fluids.
 
We would like to acknowledge useful discussions with  Ashish Lele and Arijit Bhattacharyay. We would 
also like to acknowledge computing facilities procured by DST-SERB grant no. EMR/2015/000018, and the 
Yuva cluster of CDAC-Pune.

{99}

\end{document}